\documentclass[%
 reprint,
 amsmath,amssymb,
 aps,
]{revtex4-1}

\usepackage{graphicx}% Include figure files
\usepackage{dcolumn}% Align table columns on decimal point
\usepackage{bm}% bold math
\usepackage{epstopdf}
\begin{document}

\preprint{AIP/123-QED}

\title{Double symbolic joint entropy in nonlinear dynamic complexity analysis}% Force line breaks with \\

\author{Wenpo Yao}
 \affiliation{School of Telecommunications and Information Engineering, Nanjing University of Posts and Telecommunications, Nanjing 210003, China}
\author{Jun Wang}%
 \email{wangj@njupt.edu.cn}
\affiliation{School of Geography and Biological Information, Nanjing University of Posts and Telecommunications, Nanjing 210023, China}%

\begin{abstract}
Symbolizations, the base of symbolic dynamic analysis, are classified as global static and local dynamic approaches which are combined by joint entropy in our works for nonlinear dynamic complexity analysis. Two global static methods, symbolic transformations of Wessel N. symbolic entropy and base-scale entropy, and two local ones, namely symbolizations of permutation and differential entropy, constitute four double symbolic joint entropies that have accurate complexity detections in chaotic models, logistic and Henon map series. In nonlinear dynamical analysis of different kinds of heart rate variability, heartbeats of healthy young have higher complexity than those of the healthy elderly, and congestive heart failure (CHF) patients are lowest in heartbeats' joint entropy values. Each individual symbolic entropy is improved by double symbolic joint entropy among which the combination of base-scale and differential symbolizations have best complexity analysis. Test results prove that double symbolic joint entropy is feasible in nonlinear dynamic complexity analysis.
\end{abstract}

% PACS, the Physics and Astronomy
                             % Classification Scheme.
\maketitle

%\tableofcontents

\section{Introduction}

Heart rate variability (HRV), the variation in beat-to-beat intervals represented by RR or NN interval\cite{Acharya2006}, displays irregular and non-stationary behaviors whose nonlinear dynamics provide valuable information for cardiac scientific and clinical researches\cite{Lo2015,Udhayakumar2017}. To measure its nonlinear dynamical features, some complexity parameters, such as fractal dimensions, Lyapunov exponents, geometric and entropy methods et al., are proposed\cite{Li2015,Task1996,Poincare2013}. Symbolic dynamic analysis, a kind of fast, simple and efficient method, provides rigorous ways to analyze nonlinear dynamics\cite{Hao1991}.

Symbolic time series analysis consists of symbolization and statistical analysis to the symbolic series, and it has effective applications in physiological signal analysis\cite{Guzzetti2005,Costa2008}. Symbolization involves in transforming infinite-value series into symbol sequence on basis of a given alphabet\cite{Daw2003}, so it greatly reduce demands on the data and bring convenience to series analysis\cite{Lo2015,Staniek2008}. These symbolic transformations are classified into two groups, global static and local dynamic methods\cite{Daw1998}. Symbolic transformation in works of Wessel N. et al.\cite{Wessel2007,Wessel2000} and base-scale entropy\cite{Li2006} belong to global static approaches, and symbolizations in permutation entropy\cite{Bandt2001} and differential entropy are typical local dynamic ways, and they all employ Shannon entropy for symbolic series analysis. Our objective is to take both global and local dynamical information into account to make comprehensive analysis of nonlinear dynamic complexity. There are some feasible ideas to combine the two symbolic methods such as multi-dimension theory\cite{Li2006,Bandt2001}. These attempts, however, are more appropriate to be described as compromises of, on the one hand, maintaining flexibility of global static methods and of, on the other hand, extracting sufficient local dynamic information. In order to make efficient use of the two types of symbolizations, we apply joint entropy to combine them for nonlinear dynamic complexity.

In our contributions to combine the two kinds of symbolic transformations, we conduct global static and local dynamic symbolic transformation simultaneously, and apply the two kinds of symbolizations' joint entropy to nonlinear dynamic analysis of classical nonlinear chaotic models and three kinds of real-world HRV.

\section{Symbolic transformation}

Symbolization is a course of coarse-graining or reduction, and its basic idea is to transform series $ X_{L} = \{ x_{1}, x_{2}, \ldots, x_{L}\} $ into symbolic sequence $ S_{N} = \{ s_{1}, s_{2}, \ldots, s_{N}\} $ whose element $ s_{i} $  is a finite number of symbols (letters from some alphabet). Global static methods perform symbolization according to different sequences intervals which are identified by several parameters obtained from the whole sequence, and local dynamical approaches, on other hands, take contribution of local adjacent elements' relationships to carry on symbolic representation. Both types of symbolization, targeting different types of nonlinear dynamical information, have effective applications in complexity detections.

\subsection{Wessel N. Symbolization}

To make physiology-connected symbolization which is relatively easy to interpret, Wessel N. et al. develop a four symbols context-dependent pragmatic symbols transformation\cite{Wessel2007,Wessel2000,Kurths1995}. The symbolic transformation, referring to three given levels, namely $ (1-\alpha)\mu $, $ \mu $, and $ (1+\alpha)\mu $, performs as Eq.~(\ref{eq:1})

\begin{eqnarray}
s_{i}(x_{i})=
  \left\{
       \begin{array}{lr}
          0: \mu < x_{i} \leq (1+\alpha)\mu \\
          1: (1+\alpha)\mu < x_{i} < \infty  \\
          2: (1-\alpha)\mu < x_{i} \leq \mu  \\
          3: 0< x_{i} \leq (1-\alpha)\mu
       \end{array}
  \right.
\label{eq:1}
\end{eqnarray}
where $ \mu $ is the series mean and $ \alpha $ is special controlling parameter which is recommended as from 0.03 to 0.07 according to tests and does not significantly differ resulting symbol sequences in nonlinear forecasting of cardiac arrhythmias features.

\subsection{Base-scale Symbolization}

Symbolization in base-scale entropy\cite{Li2007} is a kind of four-symbol global method, which employs multi-dimensional vector reconstruction firstly as Eq.~(\ref{eq:2}) and makes symbolic transformation in each vector.

\begin{eqnarray}
X_{m}(i)=\{x(i),x(i+\tau), \ldots, x(i+(m-1)\tau)\}
\label{eq:2}
\end{eqnarray}

In Eq.(2), $ m $ is embedding dimension and $ \tau $ is  delay time. And then base scale, the root-mean square of the differences between every two contiguous values in a vector, of each reconstructed vector is calculated as Eq.~(\ref{eq:3}).

\begin{eqnarray}
BS_{m}(i) = \sqrt{\frac{\sum^{m-1}_{j=1} [x(i+j)-x(i+j-1)]^{2}}{m-1}}
\label{eq:3}
\end{eqnarray}

The base-scale symbolic transformation goes as Eq.~(\ref{eq:4})

\begin{eqnarray}
s_{i}(x_{i})=
  \left\{
       \begin{array}{lr}
          0: \mu_{m} < x_{i} \leq \mu_{m} + \alpha \times BS_{m} \\
          1: x_{i} > \mu_{m} + \alpha \times BS_{m}  \\
          2: \mu_{m} - \alpha \times BS_{m} < x_{i} \leq \mu_{m}  \\
          3: x_{i} \leq \mu_{m} - \alpha \times BS_{m}
       \end{array}
  \right.
\label{eq:4}
\end{eqnarray}
where $ \mu_{m} $ represents the mean of m-dimension vector $ X_{m}(i) $  and $ \alpha $ describes controlling parameter which could be chosen from 0.1 to 2 accordingly. Multi-dimensional procedure brings adaptability and flexibility as well as some local dynamical information. In our works, therefore, we perform symbolic transformation on the whole time series as a vector to extract global nonlinear information.

\subsection{Permutated Symbolic Transformation}

Permutation entropy, with advantages of simplicity, fast calculation and robustness, carries on typical local dynamic symbolization\cite{Bandt2016,Zanin2012}. By comparing neighboring values and mapping time series onto symbols sequences\cite{Cao2004}, permutation entropy is a classical complexity parameter. Multi-dimensional procedure, same as base-scale entropy, is needed to transform series into symbolic sequences. Accordingly to the values' sizes, series are reorganized in for example ascending order in each reconstructed vector as $ x_{i+(j_{1}-1)\tau} \leq x_{i+(j_{2}-1)\tau} \leq \cdots \leq x_{i+(j_{m}-1)\tau}$ .

$ \pi_{j}=\{ j_{1},j_{2}, \cdots, j_{m}\} $  is a new sequence consisting of the elements' original positions, and there are $ m! $ permutations considering all possibilities. Permutation entropy is Shannon entropy of all permutations' probabilities as $ H(m)=-\sum p(\pi_{i})log_{2}p(\pi_{i}) $, where $ p(\pi_{i})\neq 0 $ .

\subsection{Symbolization in Differential Entropy}

Taking differences between adjacent elements into account, we proposed differential entropy as a dynamic complexity measure. This symbolization attributes its complexity detection to detailed local dynamic information.
The differences between current element and its forward and backward ones are $ D_{1} = x(i+\tau)-x(i) $ and $ D_{2} = x(i)-x(i-\tau) $  where $ \tau $ is the delay time.
A four-symbols transformation are carried on as Eq.~(\ref{eq:5})

\begin{eqnarray}
S_{i}(x_{i})=
  \left\{
       \begin{array}{lr}
          0: diff \geq \alpha \cdot var \\
          1: 0 \leq diff < \alpha \cdot var  \\
          2: - \alpha \cdot var < diff < 0 \\
          3: diff \geq - \alpha \cdot var
       \end{array}
  \right.
\label{eq:5}
\end{eqnarray}
where $ diff=\|D_{1}\|-\|D_{2}\| $, and $ var=\sqrt{(D_{1}^{2}+D_{2}^{2})/2} $. Parameter $ \alpha $ in could be adjusted from 0.3 to 0.6.

Code series $ C(i) $, whose formation is the next step following symbolization, is constructed by m-bit encoding of symbolic sequences, and measurements for the code series involve classical statistics and information theory, such as Shannon entropy. Taking symbols 'abc' as example, coding procedure could be $ c(i)=a*n^{2}+b*n+c $ where n should not be smaller than the amount of symbols' types, and code forms do not make significant differences to symbolic analysis. 3-bit encoding is applied in our following symbolic dynamic analysis to all symbolic sequences.

\section{Double Symbolized Dynamic Analysis}

Global static symbolic transformations flexibly select the number and size of partitions according to signals' characteristics, and local dynamic symbolizations effectively extract local detailed dynamic information. To obtain both global static and local dynamical information is the main concern of this section.

Multi-dimension\cite{Bandt2001,Li2006,Sugihara2012} vector reconstruction, a very attractive theoretical problem, is used in the base-scale entropy and permutation entropy. Through vector reconstruction, global static symbolizations is carried out in each individual vector, making base-scale entropy more adaptable to signal changes. Symbolizations of different vectors are independent from each other, and it is helpful to improve the flexibility of transformation and extract some local dynamic information. And in extreme cases, when reconstructed vector is small enough that each vector contains only 3 or even 2 elements, the global static methods are almost equivalent to local dynamic ones. In the multi-dimension method, the selection of vector length and the setting of delay factor are worthy of further and in-depth researches. Multi-dimension processing takes account of the two kinds symbolic transformations, but it is still a compromising method and cannot give fully comprehensive considerations to both sides. Another try to combine the two different symbolizations is to directly integrate the two symbolic series into new sequences. For example, the global static sequence is '0123' and local dynamic orders in '1230', and their combinated symbols are '01 12 23 30'(symbols of the global static method is in the front of each 2-bit recombination). The disadvantage of this symbolic combination is increasing amount of symbols (if symbols amounts of the two symbolizations are N and M, there are N*M symbols in the combination), but it is worth making such attempts for some unforeseen achievements.

A feasible solution is to combine the two kinds nonlinear dynamical information by joint entropy. The Shannon seminal work rationalized and initiated early efforts into information theory, which is the most influential contribution to entropy\cite{Bromiley2004}. The information contents of two (sub)systems are illustrated in Fig.1, and these relationships apply to two kinds of symbolizations as well.

\begin{figure}[htbp]
  \centering
  \includegraphics[width=6cm,height=3cm]{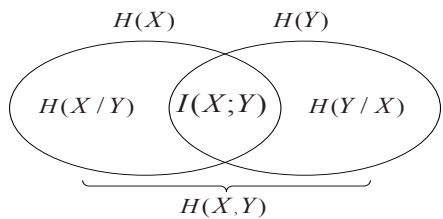}
  \caption{The relationships between information entropy of two (sub)systems}
\end{figure}

Joint entropy,  in Eq.~(\ref{eq:6}) or ~(\ref{eq:7}), is used to measure the combined amount of information.

\begin{eqnarray}
\begin{split}
H(X,Y)
&=\sum \sum p(x_{i},y_{j})I(x_{i},y_{j})\\
&=-\sum \sum p(x_{i},y_{j}) log p(x_{i},y_{j})
\end{split}
\label{eq:6}
\end{eqnarray}

\begin{eqnarray}
\begin{split}
&H(X,Y)=H(X)+H(Y)-I(X,Y)\\
&=-\sum p(x) log p(x)- \sum p(y) log p(y)+ \sum log p(x,y)
\end{split}
\label{eq:7}
\end{eqnarray}

In combination of two different symbolizations, global static and local dynamical symbolic series are obtained simultaneously as $ X_{G} $ and $ X_{L} $ whose joint entropy are calculated as $ H(X_{G},X_{L}) $. In our following double symbolic dynamic analysis, 'BS-PE JEn' describes the joint entropy of base-scale and permutated symbolization, and 'BS-DE JEn', 'WN-PE JEn' and 'WN-DE JEn' are other three combinations of base-scale and differential joint entropy, Wessel N. et al.'s and permutated joint entropy, Wessel N. et al.'s and differential joint entropy. To analyze the relationships between each individual symbolic entropy and four double joint entropies, we do not normalize all kinds of symbolic entropy methods.

\section{Double Symbolic Joint Entropy Analysis of Chaotic Models}

The four double symbolic joint entropy methods are tested by logistic and Henon map series. Delay time in the four symbolizations are set to 1. We refer to choices of controlling parameters in their original works and their performances in logistic map analysis, and set $ \alpha $  in base-scale entropy to 0.2 and differential entropy to 0.5, while value to Wessel N. symbolic entropy is 0.3.

The canonical form of logistic difference equation, $ X_{i+1}=r\cdot x_{i}(1-x_{i}) $, is attractive by virtue of its extreme simplicity\cite{May1976} and is widely applied in chaotic and nonlinear dynamical analysis. Its bifurcation diagram and chaotic detections of four double symbolic joint entropy are shown in Fig.2.

\begin{figure}[htbp]
  \centering
  % Requires \usepackage{graphicx}
  \includegraphics[width=9cm,height=2.7cm]{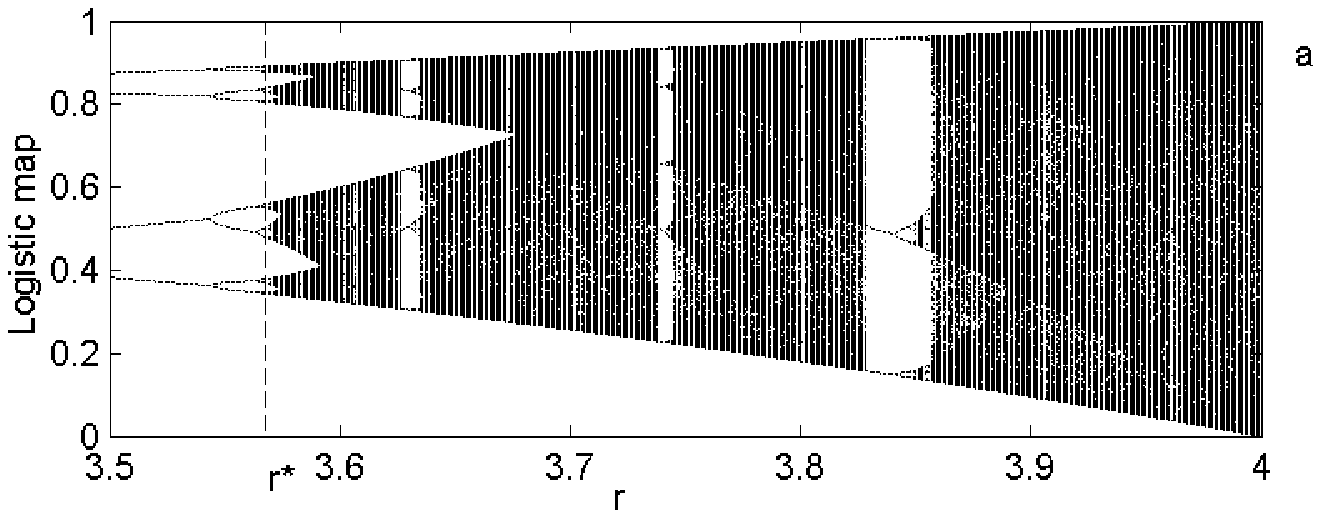}
  \includegraphics[width=9cm,height=2.7cm]{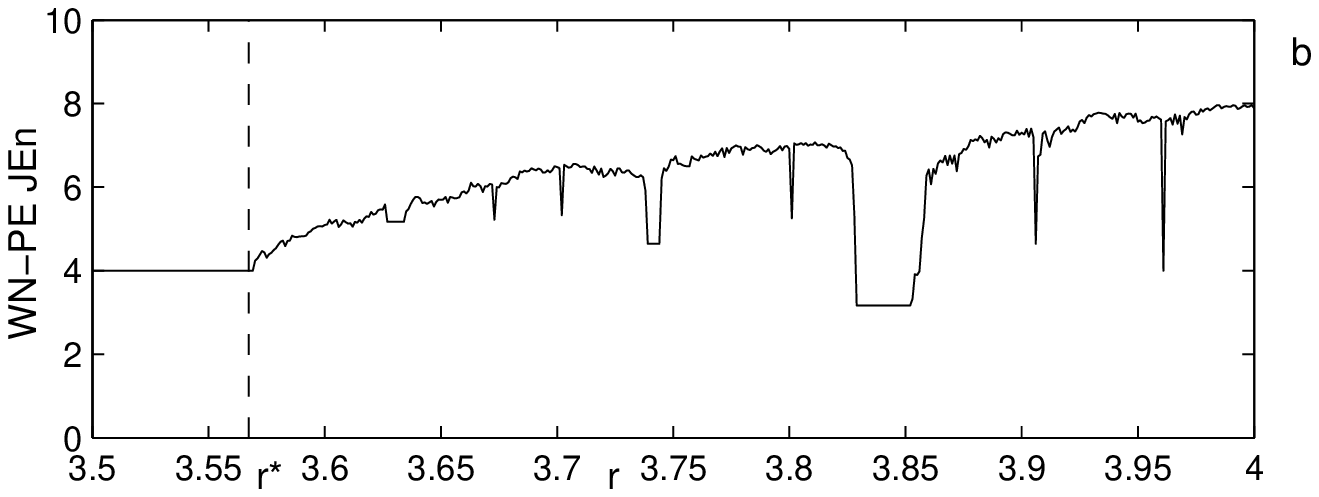}
  \includegraphics[width=9cm,height=2.7cm]{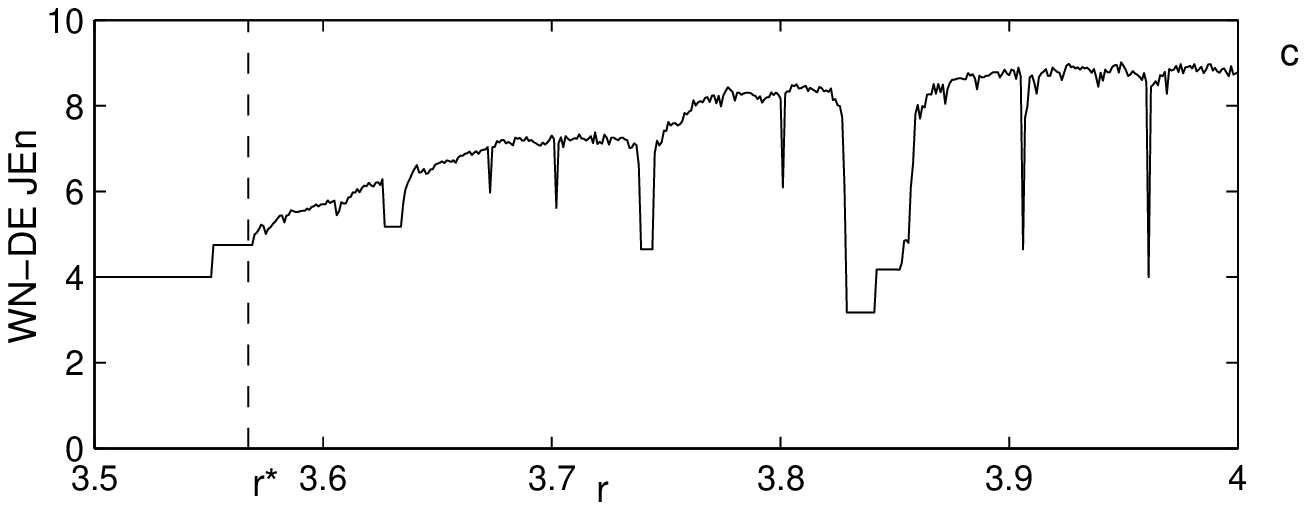}
  \includegraphics[width=9cm,height=2.7cm]{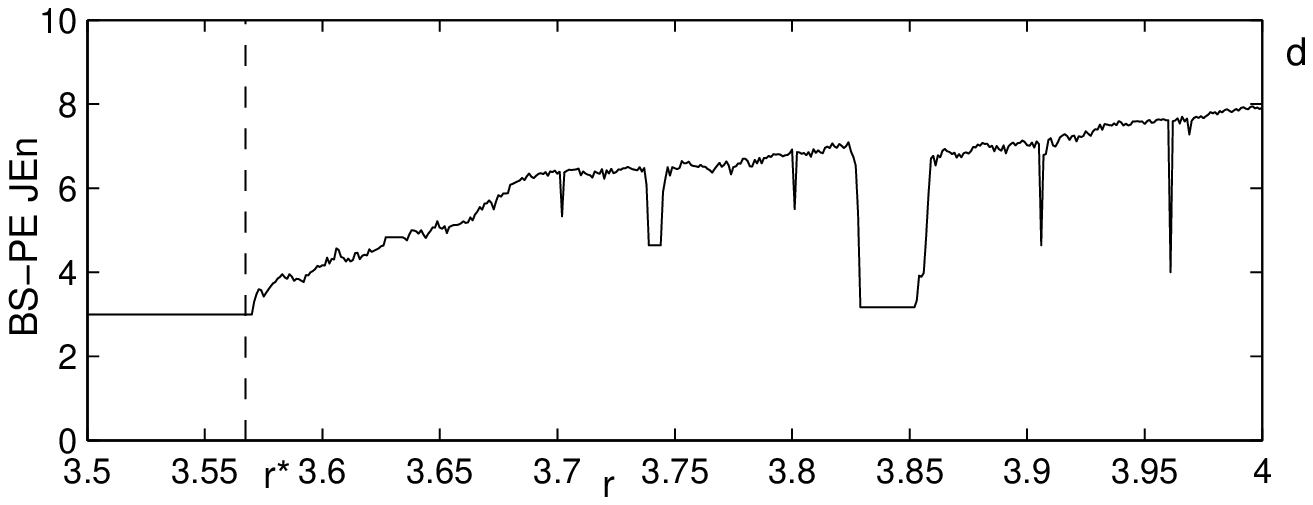}
  \includegraphics[width=9cm,height=2.7cm]{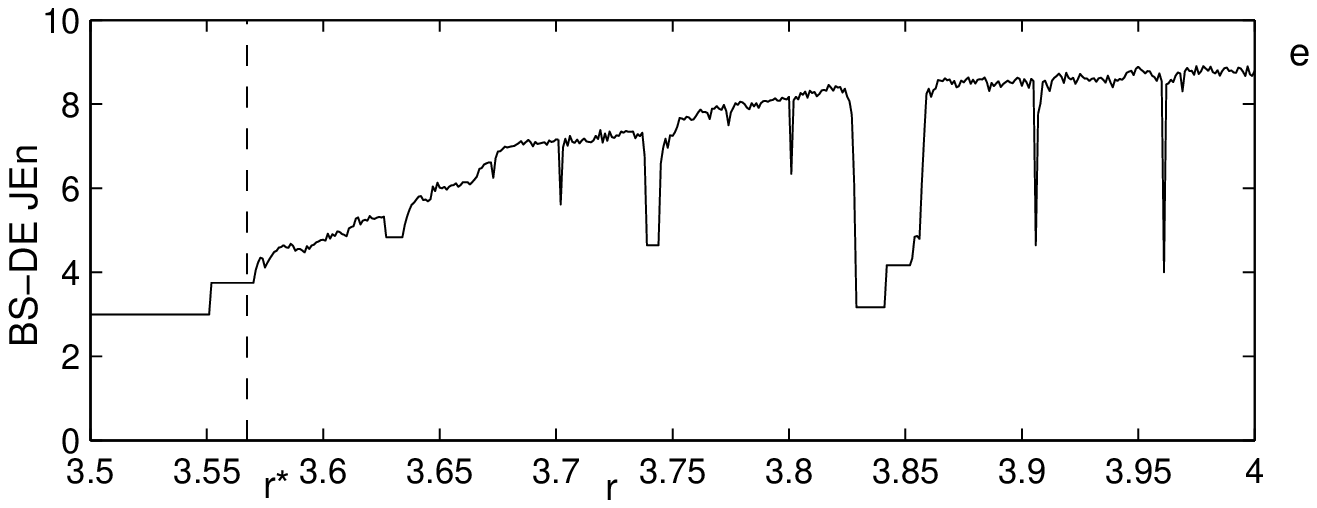}
  \caption{Logistic equations for varying control parameter from 3.5 to 4. (a) Bifurcation diagram. (b) WN-PE JEn. (c) WN-DE JEn. (d) BS-PE JEn. (e) BS-DE JEn.}
\end{figure}

Logistic map shows its chaotic characteristics when r is larger than the cut-off point r*=3.567, exactly $ \approx $ 3.569946, and its nonlinear complexity increases with the increase of r. The four combined joint entropies effectively identify chaotic behaviors at r*, and their entropy values increase with chaotic enhancement of logistic map showing in Fig.2. The double symbolized joint entropies, therefore, are effective to serve as complexity parameters.

In Henon map tests, the four symbolic combined joint entropies show their satisfied chaotic complexity detection as well. The discrete-time dynamical system is a simple mapping\cite{Henon1976} of the plane defined by  $ x_{i+1}=ry_{i}+1-1.4x^{2}_{i} $,  $ y_{i+1}=0.3rx_{i} $, where r is a controlling parameter. As r increase from 0.9 to 1, Henon system exhibits more chaotic behaviors, and its four double symbolic joint entropy are listed in Table~\ref{tab:1}.

\begin{table}
\caption{\label{tab:1}Double symbolic joint entropy in Henon map analysis.}
\begin{ruledtabular}
\begin{tabular}{ccccc}
R      &WN-PE JEn & WN-DE JEn  & BS-PE JEn & BS-DE JEn\\
\hline
0.9 	&7.7519	  &9.4581	   &7.3213	   &9.0275\\
1	   &8.4924	  &9.9976	   &7.9587	   &9.4639\\
\end{tabular}
\end{ruledtabular}
\end{table}

From Tab 1, as Henon series chaotic behaviors increase, four joint entropy methods have corresponding increase, verifying their effective nonlinear complexity detections.

\section{Double Symbolic Joint Entropy in HRV Analysis}

Three kinds of heartbeat intervals (derived from ECG) from Physionet Database\cite{Goldberger2000} are applied in our works. Firstly, 15 subjects with severe congestive heart failure (CHF), NYHA class 3-4\cite{Baim1986}, including 11 men aged 22 to 71 and 4 women aged 54 to 63. Secondly, 20 young (21 to 34 years old) and 20 elderly (68 to 85 years old) underwent 2 hours of continuous data collecting\cite{Iyengar1996} in a resting state in sinus rhythm.

We firstly apply four double symbolic joint entropy methods and individual symbolic entropy approaches to the three kinds of HRV, and analysis results are illustrated in FIG.3a and 3b. In this part, controlling parameters are all set to 0.55.

\begin{figure}[htbp]
  \centering
  \includegraphics[width=8.1cm,height=4.5cm]{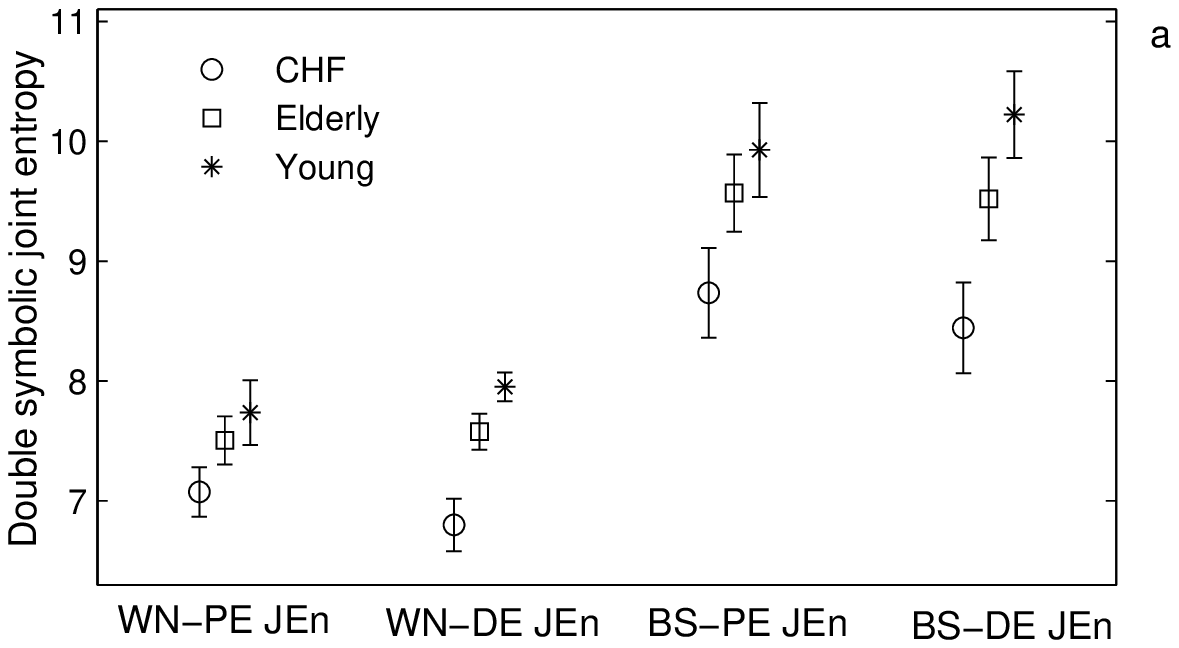}
  \includegraphics[width=8.1cm,height=4.5cm]{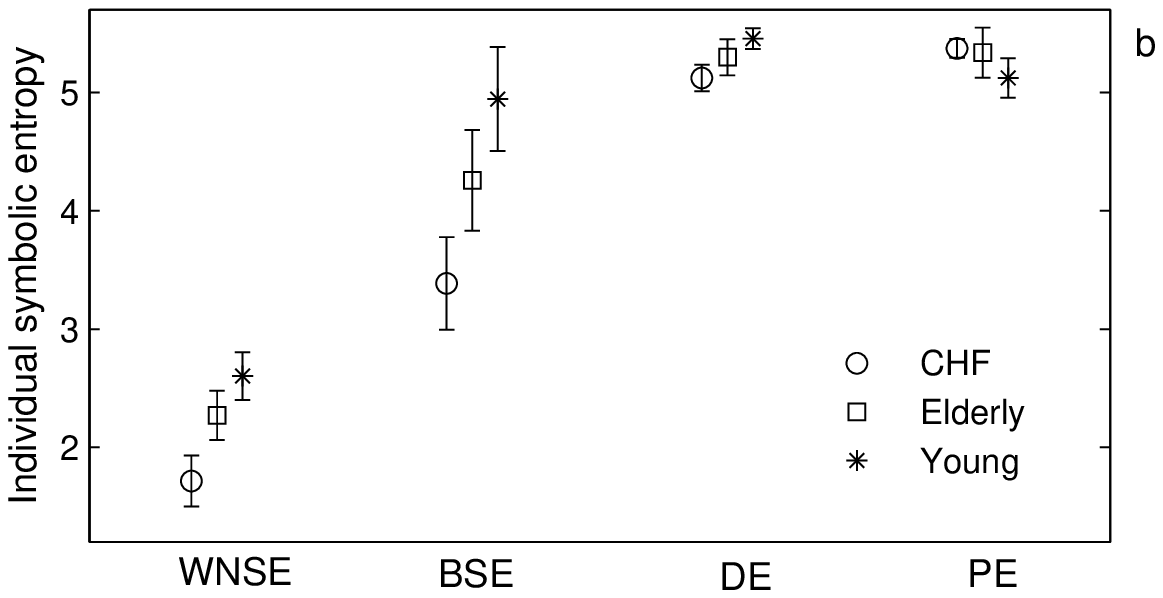}
  \caption{Symbolic entropy analysis of three groups of heartbeats. (a) Double symbolic joint entropy. (b) Individual symbolic entropy ('WNSE' represents the symbolic entropy in works of Wessel N. et al., 'BSE' accounts for the base-scale entropy, 'PE' stands for permutation entropy and 'DE' denotes differential entropy.)}
\end{figure}

From Fig.3a, the four double symbolic joint entropies distinguish the three kinds of HRV and share coincident distinctions which are consistent with the 'complexity-loss' theory of aging and disease in relevant researches\cite{Lo2015,Ho2011,Goldberger2002,Costa2005} that joint entropies of healthy young volunteers' heart rates are higher than those of healthy elderly ones, and CHF patients have the lowest joint entropy values. Healthy young subjects have better cardiac states and their heartbeats present more complex processes in nature than those of the elderly and CHF patients. The healthy elderly subjects represent slight weakness in cardiac function due to aging, and therefore their dynamic features are less than the young ones. CHF group, having profound abnormalities in cardiac function and severe damage to the cardiac control system, largely lose their HRV dynamic features, therefore their group have the lowest dynamic complexity.

T tests for the four double symbolic joint entropy analysis of three different kinds of heartbeats are carried out, and p values are listed in Table~\ref{tab:2}.

\begin{table}
\caption{\label{tab:2}p values of four double symbolic joint entropy in three kinds of heart rates analysis ('0.000' should be read as $ p<0.001 $.CHF is abbreviated to C, Elderly to E and Young to Y)}
\begin{ruledtabular}
\begin{tabular}{ccccc}
                &WN-PE JEn & WN-DE JEn  & BS-PE JEn & BS-DE JEn\\
\hline
C-E	    &0.042	        &0.011	&0.001	        &0.001\\
C-Y	    &0.000	        &0.000	&0.000	        &0.000\\
E-Y	&\textbf{0.083}	&0.001	&\textbf{0.086}	&0.009\\
\end{tabular}
\end{ruledtabular}
\end{table}

From Table~\ref{tab:2}, differences between each two kinds of heart signals' complexity extracted by WN-DE JEn and BS-DE JEn are significant ($ p<0.05 $) that the two DE-jointed methods achieve satisfied nonlinear distinctions among three groups of HRV and BS-DE joint entropy show optimal nonlinear dynamic complexity detections. WN-PE JEn and BS-PE JEn effectively separate cardiac rhythms of CHF patients and two kinds of healthy subjects while they both fail to significantly distinguish the elderly and young volunteers' heartbeats (p=0.083 and 0.086, larger than 0.05). The failures of two PE-jointed entropy in distinguishing two kinds of healthy HRV may lie in the misleadings of permutation entropy in these heart signals nonlinear analysis.

In Figure 3b, permutation entropy has different results from other three symbolic entropy that nonlinear complexities of the three kinds of HRV show oppositions. Heartbeats of CHF patients have biggest entropy of 5.372 and those of elderly persons have permutation entropy of 5.335 while complexity of healthy young subjects' heart rates, 5.121, are lowest. This paradox phenomenon, we guess, may be involved in multi-scale theory that higher complexity for certain pathologic processes, such as CHF, than for healthy dynamics in permutation entropy analysis lies in that it fails to account for the multi-scale information\cite{Ho2011,Costa2008,Costa2005,Costa2002}. Multi-scale concept is to construct coarse-grained series $ y^{\tau}_{j}=1/\tau \Sigma ^{j\tau} _{i= (j-1)\tau +1} x_{i} $, $ 1 \leq j \leq N/ \tau $ and for scale 1, $ \{ y^{1}_{j}\} $  is the original series $ \{ x_{i}\} $. The single-scale permutation entropy may be related to this inconsistency in our works . Except for permutation entropy, the other three individual symbolic entropy methods effectively distinguish the different heart signals and their results are not inconsistent with previous 'complexity-loss' theory. Independent samples t tests for other three symbolic entropy methods show that they all effectively distinguish three different kinds of heart rate variability in nonlinear dynamic analysis (CHF-Elderly p values of WNSE, BSE and DE are 0.031, 0.001 and 0.019, the Elderly-Young correspondings are 0.002, 0.024 and 0.042, and CHF-Young p values are all 0.000).

Through the above analysis, we find that double symbolic joint entropy improves nonlinear complexity extraction of individual symbolic entropy. The two DE-combined joint entropies, reducing p values of different kinds of heartbeats of each individual symbolic entropy, improve distinctions among the three groups of heart signals and more effectively identify different kinds of HRV. The two PE-combined joint entropies, having correct distinctions of the three HRV complexity shown in Fig.3a, overcome the unenviable situations of permutation entropy and effectively distinguish HRV of CHF patients and two healthy groups.

In this part, we observe impacts of data length on joint entropy analysis. Data length increases from 300 to 4500 with step size of 300 for researches on the four joint entropy analysis of HRV, and results are shown in Fig.4.

\begin{figure}[htbp]
  \centering
  \includegraphics[width=9cm,height=3cm]{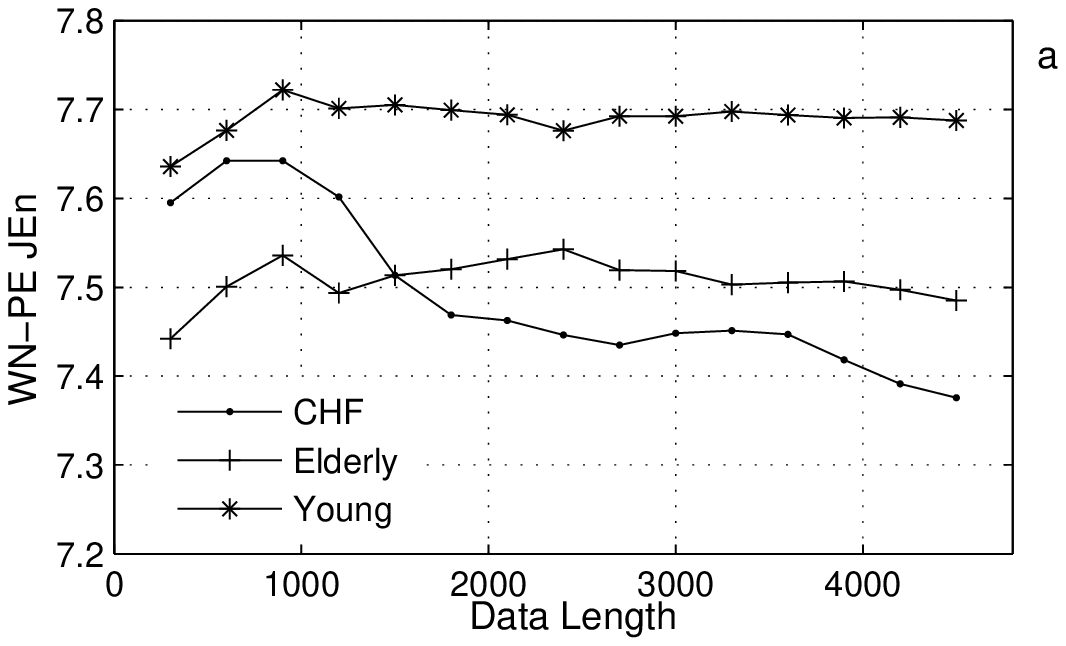}
  \includegraphics[width=9cm,height=3cm]{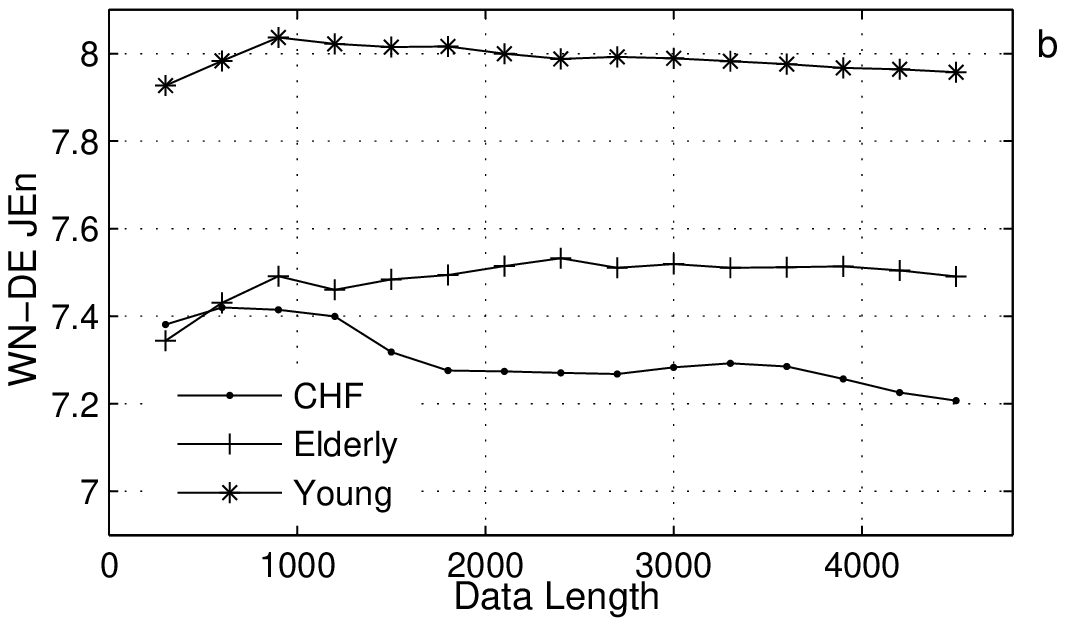}
  \includegraphics[width=9cm,height=3cm]{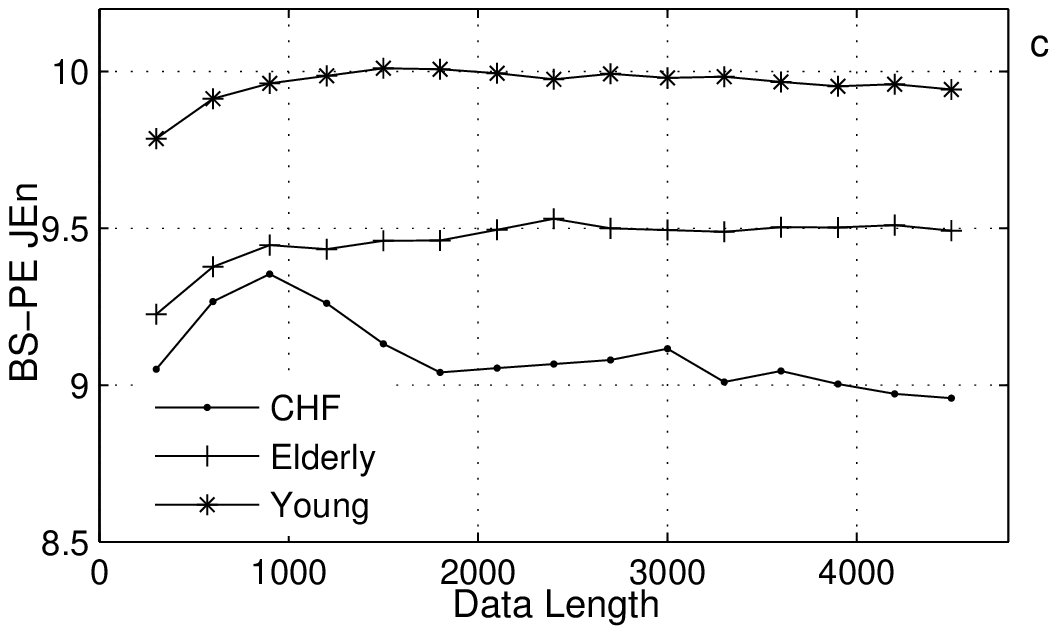}
  \includegraphics[width=9cm,height=3cm]{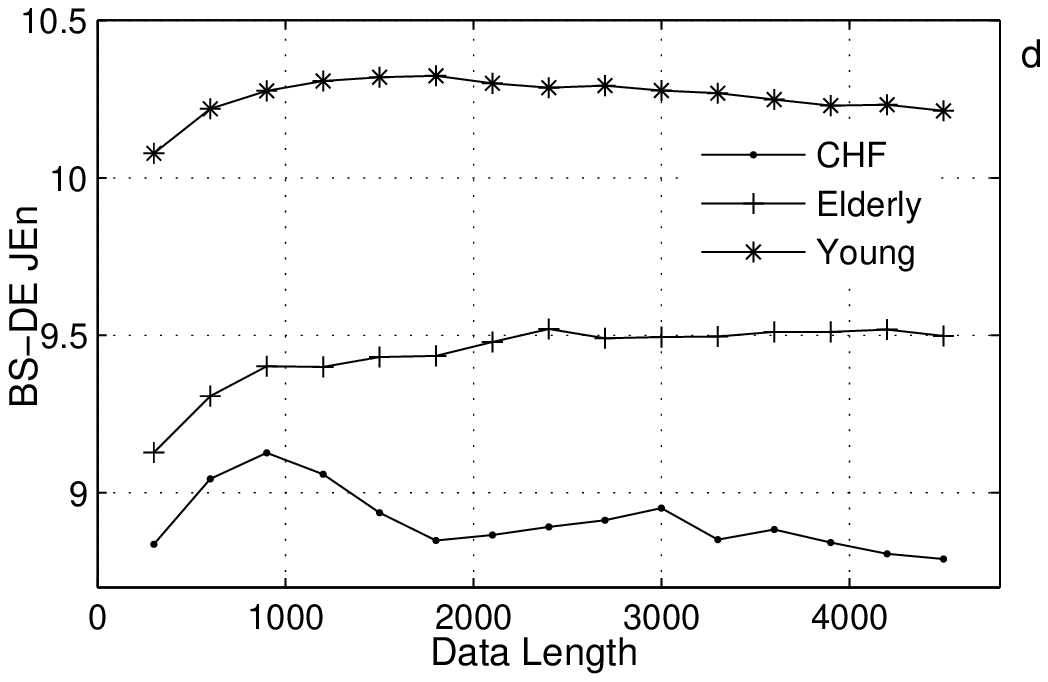}
  \caption{Joint entropies of the three kinds of HRV with the increase of data length. (a) WN-PE JEn. (b) WN-DE JEn. (c) BS-PE JEn. (d) BS-DE JEn.}
\end{figure}

Illustrated by Fig 4a and 4b, in the beginning data length of two WN-joint entropy, heartbeats of CHF patients have higher nonlinear complexity than that of the healthy elderly and their relationships change when data length become to and larger than 2000. In Fig 4c and 4d, BS-joint entropy values of three groups of HRV are consistent with normal relationships in previous analysis and become to stable when data length comes to 2000. In the Fig 4, in beginning parts entropy values of two healthy heartbeats undergo increasing trends while those of CHF patients have first-increasing and then-decreasing changes.

In the four subplots, healthy young people maintain higher entropy to the healthy elderly ones, which are not affected by the data length. The differences in charts are mainly reflecting in CHF patients entropy trends, reasons for this we suppose are that CHF patients HRV signals are in poor stability that contributes to these fluctuations.

From Fig.4, the relationships between the three HRV joint entropies change at the beginning parts and tend to converge as data lengths increase to about 2000 and larger. And we come to the point that the four double symbolic joint entropies have certain requirements for data length in HRV analysis.

\section{Discussions}

To take flexibility of global static transformations and detailed dynamic information extraction of local dynamic methods into account, we introduce joint entropy to combine characteristics of the two kinds of symbolizations.

Among the four individual symbolic transformations, there are three with controlling parameter whose adjustments play important role in nonlinear dynamic analysis. Referring to choice ranges of controlling parameters given in their original works, we make adjustment in nonlinear complexity detection of chaotic models and physiological signals accordingly. In logistic and Henon map analysis, $ \alpha $ is set 0.3 to Wessel N. symbolic entropy, 0.2 to base-scale entropy and 0.5 to differential entropy while in nonlinear complexity extraction of heartbeats $ \alpha $ are adjusted to 0.55 to all three symbolizations to achieve satisfied results. We find that the choices controlling parameter of Wessel N. symbolization, whether 0.3 in chaotic models complexity detections or 0.55 in different HRV nonlinear dynamic analysis, are not in recommended range of 0.03 to 0.07 in the original literatures\cite{Wessel2007,Wessel2000}. It seems we cannot find optimal controlling parameters for all different kinds of data. The reasons, we guess, account for this lie in differences of structural information or of dynamical complexity in different types of signals, so the parameters should be adjusted accordingly. And it need to be validated that whether our findings apply to other nonlinear signals.

Both global static and local dynamic symbolizations contain irreplaceable dynamic information about series, and there is not too much redundant information in two types of symbolic transformations. Taking heartbeats of the first CHF patient 'chf01' as an example, four combined symbolic joint entropies are 7.0001, 6.9738, 8.8819 and 8.8555 which are approximately equal to the sum of each two individual entropy which are 1.6959 to WNSE, 3.5776 to BSE, 5.2779 to DE and 5.3042 to PE. The same results are true for those of the healthy elderly persons and CHF patients. Joint entropy values are close to the sum of each symbolization entropy, proving that the two symbolic transformation approaches extract series nonlinear dynamic information from different perspectives, and degree of repetition of the two different symbolizations is very low.

\section{Conclusions}

The above analysis and tests show that it is feasible and effective to use joint entropy of global static and local dynamic symbolizations for series nonlinear dynamic complexity detection. Double symbolic joint entropy is beneficial to improve final nonlinear complexity detections of each individual symbolic entropy in our nonlinear analysis of heart rate variability. And the 'complexity-loss' theory of aging and disease is validated in our contributions.

\section{Acknowledgments}

The work is supported by Project supported by the National Natural Science Foundation of China
(Grant Nos. 61271082, 61401518, 81201161), Jiangsu Provincial Key R \& D Program (Social Development) (Grant No.BE2015700), the Natural Science Foundation of Jiangsu Province (Grant No. BK20141432), Natural Science Research Major Programmer in Universities of Jiangsu Province (Grant No.16KJA310002), Postgraduate Research \& Practice Innovation Program of Jiangsu Province (KYCX17-0788).

\nocite{*}

\bibliography{mybibfile}% Produces the bibliography via BibTeX.

\end{document}